\input{aipcheck}

\documentclass[final
    ,final            % use final for the camera ready runs
  ]
  {aipproc}

\layoutstyle{6x9}
\usepackage{graphicx}
\usepackage{amssymb,amsmath}

\begin{document}

\title{Interacting viscous matter with a dark energy fluid}


\author{Arturo Avelino}{
  address={Departamento de F\'isica, DCI, Campus Le\'on,
Universidad de Guanajuato, \\
C\'odigo Postal 37150, Le\'on, Guanajuato, Mexico.} }

\begin{abstract}
We study a cosmological model composed of a dark energy fluid
interacting with a viscous matter fluid in a spatially flat Universe. 
The matter component represents the baryon and dark matter and it is taken
into account, through a bulk viscosity, the irreversible process that the
matter fluid undergoes because of the accelerated expansion of the
universe. The bulk viscous coefficient is assumed to be proportional to the
Hubble parameter. The radiation component is also taken
into account in the model. The model is constrained using the type Ia
supernova observations, the shift parameter of the CMB, the acoustic peak
of the BAO and the Hubble expansion rate, to constrain the values of the
barotropic index of dark energy and the bulk viscous coefficient.
It is found that the bulk viscosity is constrained to be negligible
(around zero) from the observations and that the
barotropic index for the dark energy to be negative and close to zero
too, indicating a phantom energy.

\end{abstract}

\maketitle

%%%%%%%%%%%%%%%%%%%%%%%%%%%%%%%%%%%%%%%%%%%%
%% MAINMATTER
%%%%%%%%%%%%%%%%%%%%%%%%%%%%%%%%%%%%%%%%%%%%

\section{Introduction}

In the last years, the type Ia supernovae (SNe Ia) observations have
given a strong evidence of a present accelerated expansion epoch of
the Universe  (see for instance
\cite{Riess:1998cb,Perlmutter:1998np,Amanullah:2010vv} and references
therein).

Several models have been proposed to explain this recent acceleration,
one of the most successful one is the so-called $\Lambda$ Cold Dark
Matter ($\Lambda$CDM) that proposes the existence of a new kind of
component in the Universe called ``dark energy'' with a behavior of a
cosmological constant and that constitutes  $\sim 73\%$ of the
total content of matter-energy in the Universe today, in addition
to a dark matter component filling the Universe in a $\sim 23\%$
\cite{Amanullah:2010vv}.

However, this model faces several strong problems,
one of them is the huge discrepancy between its predicted and observed
value for the dark energy density (of about 120 orders of magnitude)
\cite{WeinbergCosmoConstProblem,Padmanabhan:2002ji,Carroll:2000fy},
another one is the so-called
the ``\emph{cosmic coincidence problem}'': the model predicts that we
are living in a moment when the matter density in the universe is of
the \textit{same order} of magnitude than the dark energy density
\cite{Steinhardt:1999nw}.

On the other hand, cosmological models with interacting dark
components have been studied by several authors, because it is
expected that the two \textit{dominant} components (dark energy and
matter) interact each other in some way. It has been found that these
models are
promising mechanisms to solve the $\Lambda$CDM problems (see
\cite{ChimentoJakubiPavon2000,Kremer:2011cd} and references therein).

%--------

In addition, it has been known since several years ago before
 the discovery of the present acceleration that a bulk viscous fluid
can produce an accelerating cosmology (although it was originally
proposed in the context of an inflationary period in the early
universe)
\cite{HellerKlimek1975,Barrow1986,Visco-Padmanabhan1987,Visco-Gron1990,
Visco-Maartens1995,Visco-Zimdahl1996}.

So, it is natural to think of the bulk viscous pressure as one of
the possible mechanism that can accelerate the universe today (see
for instance
\cite{Cataldo:2005qh,Colistete:2007xi,Avelino:2008ph,
Avelino:2010pb,
Visco-RicaldiVeltenZimdahl2010,Visco-AMontielNBreton2011}).
However, this idea faces the problem of that it is necessary to
propose a viable mechanism for the origin of the bulk
viscosity, although in this sense some proposals have been already
suggested \cite{Zimdahl:1999tn,Mathews:2008hk}.

In the present work, following the idea of Kremer et al (2011)
\cite{Kremer:2011cd} and using the SNe Ia, the shift parameter $R$ of
the cosmic microwave background radiation (CMB), the baryon acoustic
oscillation (BAO) and the Hubble expansion rate $H(z)$ data, we test
an interacting dark sector model taking into account dissipative
process through a bulk viscosity in the matter (baryon and dark
matter) component, where the interaction term is written in terms of
the barotropic index of the dark energy fluid.

In section 1 we present the characteristics of the
model and the main equations, in section 2 we
explain the cosmological probes used to constrain the model and in
section 3 we give our conclusions.

%-################################################

\section{Interacting dark fluids with bulk
viscosity}\label{SectionTheory}

We study a cosmological model in a spatially  flat FRW universe,
composed of three fluids: radiation, matter and a dark energy fluid
components. It is assumed the matter component as a pressureless
fluid, representing the baryon and dark matter, with a bulk
viscosity and interacting with the dark energy fluid.

The Friedmann constraint and the conservation equations can be
written as 
\begin{align}
H^2 &= \frac{8 \pi G}{3}\left( \rho_{\rm r} + \rho_{\rm
m} +
\rho_{\rm de} \right), \label{ConstrainFriedmann} \\
0 &= \dot{\rho}_{\rm r} + 4H\rho_{\rm r}, \label{ConsEqRadiation}\\
0 &= \dot{\rho}_{\rm m} + \dot{\rho}_{\rm de} + 3H \left(\rho_{\rm
m} + \rho_{\rm de} + p_{\rm m} + p_{\rm de} -3H\zeta \right),
\label{ConserEquation_m-de}
\end{align}

\noindent where $(\rho_{\rm r}, \rho_{\rm m}, \rho_{\rm
de})$ are the densities of the radiation, matter and dark
fluid components respectively, and $(p_{\rm r}, p_{\rm
m}, p_{\rm de})$ are their corresponding pressures. 
The equation (\ref{ConserEquation_m-de}) arises from assuming the
interaction between the matter and dark fluid components. 
The term $-3H\zeta$ corresponds to the bulk viscous pressure of
the matter fluid, where $\zeta$ is the bulk viscous
coefficient.

The immediate solution of the conservation equation
(\ref{ConsEqRadiation}) is
\begin{equation}
\rho_{\rm r}(a) =\rho_{\rm r0}/a^4, \label{DensityRadiation}
\end{equation}

\noindent where $a$ is the scale factor and the subscript zero labels
the present values for the densities.

On the other hand, following the idea of Kremer and Sobreiro
\cite{Kremer:2011cd}, 
%the interacting term between the matter and dark energy components is
%expressed in terms of the barotropic indexes of the fluids present in the
%model.
%The  idea is that the nature of the dark energy (established through
%its barotropic index) implies the nature of its
%corresponding interaction term with the matter component.
the conservation equation (\ref{ConserEquation_m-de}) can be
decoupled  as
\begin{align}
\dot{\rho}_{\rm m} + 3H\gamma^e_{\rm m} \rho_{\rm m} =0,
\label{ConsEqDM}\\
\dot{\rho}_{\rm de} + 3H\gamma^e_{\rm de} \rho_{\rm de} =0,
\label{ConsEqDE}
\end{align}

\noindent where it was defined the effective barotropic indexes
$\gamma^e_{\rm
m}$ and $\gamma^e_{\rm de}$ so that they are related as
\begin{equation}\label{EqGralIndicesBarotropicos}
\gamma^e_{\rm m} = \gamma_{\rm m} + \frac{\gamma_{\rm
de}-\gamma^e_{\rm
de}}{r} - \frac{3H \zeta}{\rho_{\rm m}},
\end{equation}

\noindent with $r \equiv \rho_{\rm m}/\rho_{\rm de}$ corresponds to
the ratio between the matter to  dark energy densities and $p_i =
(\gamma_i-1)\rho_i$  with $\gamma_i$ is the usual constant barotropic
indexes of the equation of state.

We consider a bulk viscous coefficient $\zeta$ proportional to the total
matter-energy density $\rho_{\rm t} = \rho_{\rm r} + \rho_{\rm m} +
\rho_{\rm de}$, as
\begin{equation}\label{ViscosityDefinition}
\zeta = \frac{\zeta_0}{\sqrt{24\pi G}} \rho^{1/2}_{\rm t},
\end{equation}

\noindent with $\zeta_0$ a dimensionless constant. This parametrization
corresponds to a bulk viscosity proportional to the expansion rate of the
Universe, i.e., to the Hubble parameter [see eq.
(\ref{ConstrainFriedmann})].

Following  \cite{Kremer:2011cd} and \cite{Chimento:2007yt}, we
assume that the effective barotropic index for the dark energy
is given as
\begin{equation}\label{GammaFr}
\gamma^e_{\rm de} = \gamma_{\rm de} + \zeta_0.
\end{equation}

So, using (\ref{EqGralIndicesBarotropicos}) and
(\ref{GammaFr}), the
effective conservation equations
(\ref{ConsEqDM}) and (\ref{ConsEqDE}) can be rewritten as
\begin{align}\label{EqConservationEffective3}
\dot{\rho}_{\rm m}  + 3H\gamma_{\rm m}\rho_{\rm m} &= 3H \rho_{\rm
de}\zeta_0 + 9H^2\zeta, \\
\dot{\rho}_{\rm de}  + 3H\gamma_{\rm de}\rho_{\rm de} & = -3H\rho_{\rm
de}\zeta_0, \label{EqConservationEffective4}
\end{align}
\noindent where it can be identified the interacting term $Q \equiv 3H
\rho_{\rm de}\zeta_0$.

Using the expression (\ref{GammaFr}), the solution of the
conservation equation (\ref{ConsEqDE}) becomes 
\begin{equation}\label{DensityDE2}
\rho_{\rm de}(a) =\rho_{\rm de0}/a^{3(\gamma_{\rm de} +\zeta_0)}.
\end{equation}

On the other hand, with the eqs. (\ref{ViscosityDefinition}) and
(\ref{GammaFr}) we can express the equation
(\ref{EqGralIndicesBarotropicos}) as
\begin{equation}\label{Gammas2}
\gamma^e_{\rm m} = \gamma_{\rm m} - \frac{\zeta_0}{\rho_{\rm m}}
\left(\rho_{\rm de}+\frac{3H^2}{8 \pi G} \right),
\end{equation}

\noindent that using the Friedmann constraint
(\ref{ConstrainFriedmann}) we arrive to
\begin{equation}\label{Gammas3}
\gamma^e_{\rm m} = \gamma_{\rm m} - \frac{\zeta_0}{\rho_{\rm m}}
\left(\rho_{\rm r} + \rho_{\rm m} + 2\rho_{\rm de} \right).
\end{equation}

Inserting the eqs. (\ref{DensityRadiation})  and
(\ref{DensityDE2}) at (\ref{Gammas3}) we obtain
\begin{equation}\label{Gammas4}
\gamma^e_{\rm m} = \gamma_{\rm m} - \frac{\zeta_0}{\rho_{\rm m}}
\left( \frac{\rho_{\rm r0}}{a^4}  +
2\frac{\rho_{\rm de0}}{a^{3(\gamma_{\rm de} + \zeta_0)}} + \rho_{\rm
m} \right).
\end{equation}

With this, the eq. (\ref{ConsEqDE}) for the matter density
becomes
\begin{equation}\label{ConsEqDM-2}
\dot{\rho}_{\rm m} + 3H \gamma_{\rm m} \rho_{\rm m} - 3H\zeta
\left( \frac{\rho_{\rm r0}}{a^4}  +
2\frac{\rho_{\rm de0}}{a^{3(\gamma_{\rm de} + \zeta_0)}} + \rho_{\rm
m} \right) =0.
\end{equation}

\noindent Dividing to (\ref{ConsEqDM-2}) by the present critical
density $\rho_{\rm crit}^0 \equiv 3H^2_0/ (8 \pi G)$ with $H_0$ the
Hubble constant,  and defining the dimensionless parameter
densities $\Omega_{i0} \equiv \rho_{i0}/\rho^0_{\rm crit}$, the eq.
(\ref{ConsEqDM-2}) becomes

\begin{equation}\label{ConsEqDM-OmegaDMa}
\frac{d \hat{\Omega}_{\rm m}}{da} + \frac{3}{a}
\left[\hat{\Omega}_{\rm m} (\gamma_{\rm m}-\zeta_0) - \zeta_0
\left( \frac{\Omega_{\rm r0}}{a^4} +  \frac{2\Omega_{\rm
de0}}{a^{3(\gamma_{\rm de}+\zeta_0)}} \right) \right] =0,
\end{equation}

\noindent or in terms of the redshift $z$ with the help of the
relation $a=1/(1+z)$,
\begin{equation}\label{ConsEqDM-OmegaDMz}
(1+z)\frac{d \hat{\Omega}_{\rm m}}{dz} - 3
\left[\hat{\Omega}_{\rm m} (\gamma_{\rm m}-\zeta_0) - \zeta_0
\left( \Omega_{\rm r0} (1+z)^4 + 2\Omega_{\rm de0} (1+z)^{3 ( \gamma_{\rm
de} + \zeta_0)} \right) \right] =0,
\end{equation}

\noindent where it has been defined $\hat{\Omega}_{\rm m} \equiv
\rho_{\rm m}/\rho^0_{\rm crit}$.
The analytical solution of this ordinary differential equation (ODE)
for $\hat{\Omega}_{\rm m}(z)$ is 

\begin{align}\label{SolutionOmegaM}
\hat{\Omega}_{\rm m}(z) = & [ (1+z)^{-3\zeta_0} [2
(1+z)^{3(\gamma_{\rm de} + 2 \zeta_0)} \zeta_0 (4-3\gamma_{\rm m} +
3\zeta_0) (\Omega_{\rm m0} + \Omega_{\rm r0} -1) -\\
& -3(1+z)^{4+3\zeta_0} \zeta_0 (\gamma_{\rm de} - \gamma_{\rm m} +
2\zeta_0) \Omega_{\rm r0} + \nonumber \\
& (1+z)^{3 \gamma_{\rm m}} ( (4 - 3
\gamma_{\rm m} + 3 \zeta_0) (2 \zeta_0 + (\gamma_{\rm de} -
\gamma_{\rm m}) \Omega_{\rm m0}) + \nonumber \\
& + (3\gamma_{\rm de} + 3\gamma_{\rm
m} - 8) \zeta_0 \Omega_{\rm r0} ) ] ] / \nonumber \\
& \left( (\gamma_{\rm de} - \gamma_{\rm m} + 2 \zeta_0)
(4-3\gamma_{\rm m} + 3 \zeta_0) \right). \nonumber
\end{align}

So, using the solution (\ref{SolutionOmegaM}), the Hubble parameter
(\ref{ConstrainFriedmann}) can be written as
\begin{equation}\label{HubbleParameterE}
E^2(z)  = \Omega_{\rm r0} (1+z)^4  + \Omega_{\rm de0} (1+z)^{3 (
\gamma_{\rm de} + \zeta_0)} + \hat{\Omega}_{\rm m}(z),
\end{equation}

\noindent where $E(z) \equiv H(z)/H_0$. 
In the following we will assume $\gamma_{\rm m} = 0$, i.e., the matter
as a pressureless fluid.

\section{Cosmological probes}
\label{SectionProbes}

We compare the model with the following cosmological probes that
measure the expansion history of the Universe, to constrain the values
of $(\zeta_0, \gamma_{\rm de})$.

\subsubsection{Type Ia Supernovae}
We use the type Ia supernovae (SNe Ia) of the ``Union2'' data set
(2010) from the Supernova Cosmology Project (SCP) composed of 557 SNe
Ia \cite{Amanullah:2010vv}.
The luminosity distance $d_L$ in a spatially flat Universe is defined
as

\begin{equation}
d_L(z,\zeta_0, \gamma_{\rm de}, H_0) = \frac{c(1+z)}{H_0} \int_0^z
\frac{dz'}{E(z',\zeta_0, \gamma_{\rm de})},
\end{equation}

\noindent where ``$c$'' corresponds to the speed of light in units of
km/sec. The theoretical distance moduli $\mu^t$ for the k-th supernova
at a distance $z_k$ given by

\begin{equation}
\mu^t(z,\zeta_0, \gamma_{\rm de}, H_0) = 5 \log \left[
\frac{d_L(z,\zeta_0, \gamma_{\rm de}, H_0) }{\rm Mpc} \right] + 25.
\end{equation}

So, the $\chi^2$ function is defined as

\begin{equation}
\chi^2_{\rm SNe}(\zeta_0, \gamma_{\rm de}, H_0) \equiv \sum_{k=1}^n
\left( \frac{\mu^t(z_k,\zeta_0, \gamma_{\rm de}, H_0)-
\mu_k}{\sigmạ_k} \right)^2,
\end{equation}

\noindent where $\mu_k$ is the observed distance moduli of the k-th
supernova, with a standard deviation of $\sigmạ_k$ in its
measurement, and $n= 557$.

\subsubsection{Cosmic Microwave Background Radiation}
We use the WMAP 7-years distance priors release shown in table 9 of
\cite{Komatsu:2010fb}, composed of the shift parameter $R$, the
acoustic scale $l_A$ and the redshift of decoupling $z_*$.

The shift parameter $R$ is defined as 
\begin{equation}
R = H_0 \frac{\sqrt{\Omega_{\rm m0}}}{c} (1+z_*) D_A(z_*),
\end{equation}
\noindent where $D_A$ is the proper angular diameter distance given
by (for a spatially flat Universe)
\begin{equation}
D_A(z)= \frac{c}{(1+z)H_0} \int^z_0 \frac{dz'}{E(z',\zeta_0,
\gamma_{\rm de})}.
\end{equation}

With $R$ we can defined a $\chi^2$ function as
\begin{equation}\label{Chi2RCMB}
\chi^2_{\rm R-CMB}(\zeta_0, \gamma_{\rm de}, H_0) \equiv \left(
\frac{R(\zeta_0, \gamma_{\rm de}, H_0) - R_{\rm obs}}{\sigma_{R}}
\right)^2,
\end{equation}
\noindent where $R_{\rm obs} = 1.725$ is the ``observed'' value of the
shift parameter and $\sigma_{R}=0.018$ the standard deviation of the
measurement (cf. table 9 of \cite{Komatsu:2010fb}).

The acoustic scale $l_A$ is defined as
\begin{equation}
l_A \equiv (1+z_*)\frac{\pi D_A(z_*)}{r_s(z_*)},
\end{equation}
\noindent where $r_s(z_*)$ corresponds to the comoving sound horizon
at the decoupling epoch of photons, $z_*$, given by
\begin{equation}
r_s(z)=\frac{c}{\sqrt{3}} \int_0^{1/(1+z)} \frac{da}{a^2 H(a)
\sqrt{1+(3 \Omega_{\rm b0} / 4 \Omega_{\gamma 0})a}},
\end{equation}
\noindent where we use $\Omega_{\gamma 0} = 2.469 \times 10^{-5}
h^{-2}$ the radiation, and $\Omega_{\rm b0} = 0.02255 h^{-2}$ the baryon
matter component, as reported by Komatsu et al. 2010
\cite{Komatsu:2010fb}.
For $z_*$ we use the fitting formula proposed by Hu and Sugiyama
\cite{Hu:1995en}
\begin{equation}
z_*=1048 \left[ 1+0.00124 (\Omega_{\rm b0} h^2)^{-0.738} \right] \left[ 1
+ g_1 (\Omega_{\rm m0} h^2)^{g_2} \right],
\end{equation}
\noindent where 
\begin{equation}
g_1 = \frac{0.0783
(\Omega_{\rm b0}h^2)^{-0.238}}{1+39.5(\Omega_{\rm b0}h^2)^{0.763}}, \; \;
\; \; \; g_2= \frac{0.560}{1+21.1(\Omega_{\rm b0} h^2)^{1.81}}.
\end{equation}

The $\chi^2$ function using the three values $(l_A, R, z_*)$ is
defined as
\begin{equation}\label{chi2CMB3p}
\chi^2_{\rm CMB}(\zeta_0, \gamma_{\rm de}, H_0) =  \sum_{i,j=1}^3 (x_i
- d_i)(C^{-1})_{ij}(x_j - d_j),
\end{equation}
\noindent where $x_i \equiv (l_A, R, z_*) $ are the predicted values
by the model and $d_i \equiv (l_A = 302.09, R=1.725, z_*=1091.3) $
are the observed ones and $C^{-1}_{ij}$ is the inverse covariance
matrix \cite{Komatsu:2010fb}

%The subscript ``3'' at eq. (\ref{chi2CMB3p}) indicates that it is
%used the three $(l_A, R, z_*)$ points as a probe to test the model.

\begin{equation}
C^{-1} =
\begin{pmatrix}
2.305 & 29.698 & -1.333 \\
29.698 & 6825.27 & -113.180 \\
1.333 & 113.180 & 3.414
\end{pmatrix}.
\end{equation}

\subsubsection{Baryon Acoustic Oscillations}

We use the baryon acoustic oscillation (BAO) data from the SDSS
7-years release \cite{Percival:2009xn}. The distance ratio
$d_z$ at $z=0.275$ is defined as
\begin{equation}
d_{0.275} \equiv \frac{r_s(z_d)}{D_V(0.275)},
\end{equation}
\noindent where $z_d$ is the redshift at the baryon drag epoch
computed from the following fitting formula
\cite{Eisenstein:1997ik}
\begin{align}
z_d &= 1291 \frac{(\Omega_{m0} h^2)^{0.251}}{1+0.659(\Omega_{m0}
h^2)^{0.828}} \left[ 1 + b_1 (\Omega_{m0} h^2)^{b_2} \right], \\
b_1 &= 0.313 (\Omega_{m0} h^2)^{-0.419} \left[1 + 0.607 (\Omega_{m0}
h^2)^{0.674} \right], \\
b_2 &= 0.238 (\Omega_{m0} h^2)^{0.223}.
\end{align}

For a flat Universe, $D_V(z)$ is defined as

\begin{equation}
D_V(z) = c \left[ \left( \int_0^z \frac{dz'}{H(z')} \right)^2
\frac{z}{H(z)} \right]^{1/3},
\end{equation}
\noindent contains the information of the visual distortion of a
spherical object due the non Euclidianity of the FRW spacetime.

The $d_{0.275}$ contains the information of the other two pivots, 
$d_{0.2}$ and $d_{0.35}$ usually used for other authors, with a
precision of $0.04\%$ \cite{Percival:2009xn}.

The $\chi^2$ function for BAO is defined as
\begin{equation}\label{Chi2BAO}
\chi^2_{\rm BAO}(\zeta_0, \gamma_{\rm de}, H_0) \equiv \left(
\frac{d_{0.275} - d_{0.275}^{\rm obs}}{\sigma_{d}} \right)^2,
\end{equation}
\noindent where $d_{0.275}^{\rm obs} = 0.139$ is the ``observed''
value and $\sigma_{d} = 0.0037$ the standard deviation
of the measurement \cite{Percival:2009xn}.

\subsubsection{Hubble expansion rate}
For the Hubble parameter we use 13 available data, 11 comes from the
table 2 of Stern et al. (2010) \cite{Stern:2009ep} and
the 2 following data from Gaztanaga et al. 2010
\cite{Gaztanaga:2008xz}: $H(z=0.24)=79.69 \pm 2.32$ and $H(z=0.43)=
86.45 \pm 3.27$ km/s/Mpc. 
For the present value of the Hubble parameter we take that reported
by Riess et al 2011 \cite{Riess:2011yx} $H(z=0) \equiv H_0
= 73.8 \pm 2.4$ km/s/Mpc.
The $\chi^2$ function is defined as
\begin{equation}\label{Chi2Hz}
\chi^2_{\rm H}(\zeta_0, \gamma_{\rm de}, H_0) = \sum_i^{13} \left(
\frac{H(z_i, \zeta_0, \gamma_{\rm de}) - H_i^{\rm obs}}{\sigma_{H} }
\right)^2,
\end{equation}

\noindent where $H(z_i)$ is the theoretical value predicted by the
model and $H_i^{\rm obs}$ is the observed value.

%----------------------------------------

\begin{figure}
\includegraphics[width=14cm]{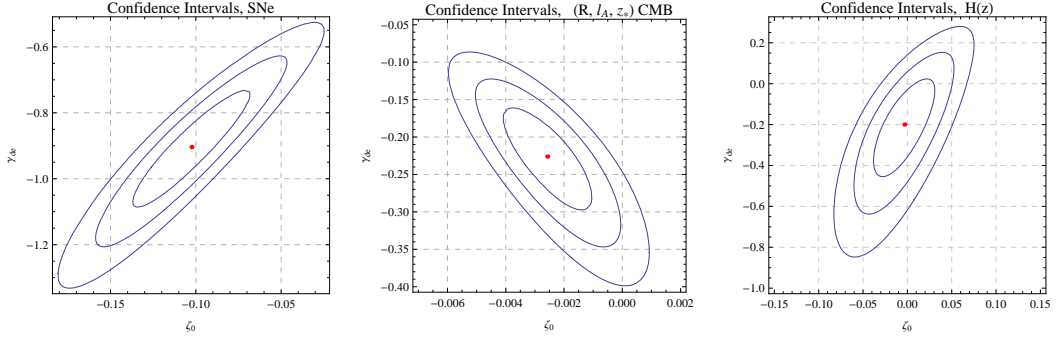}
\caption{Confidence intervals for $(\zeta_0, \gamma_{\rm de})$. The
left panel corresponds to the constraint using the SNe Ia ``Union2''
data set. The central panel, using the $(R, l_A, z_*)$ data from the
CMB and the left panel to the Hubble parameter $H(z)$. It was assumed
a value of $H_0=73.8$ km/s/Mpc as suggested by \cite{Riess:2011yx}.
The best estimated values and the $\chi^2$ minimum values are shown
in table \ref{TableBestEstimated}. The contours correspond to 63.8\%, 95\%
and 99\% of confidence level.}
\label{PlotCISNeU2CMB3pHz}
\end{figure}

\begin{figure}
\includegraphics[width=14.5cm]{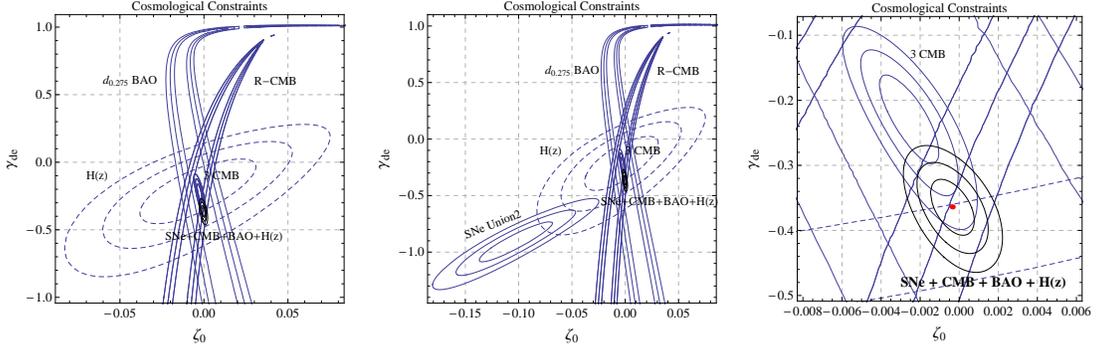}
\caption{Confidence intervals (CI) together for $(\zeta_0, \gamma_{\rm
de})$. They correspond to the constraints when it is used the SNe Ia
dataset, the $d_{0.275}$ BAO probe (see eq. [\ref{Chi2BAO}]), the shift
parameter $R$ of the CMB (see eq. [\ref{Chi2RCMB}]), the three
$(R, l_A, z_*)$ data from the CMB (see eq. [\ref{chi2CMB3p}]), the Hubble
parameter $H(z)$ (see eq. [\ref{Chi2Hz}]),  and the joint ``SNe + CMB + BAO
+ $H(z)$'' (black contours).
It was assumed a value of $H_0=73.8$ km/s/Mpc as suggested by
\cite{Riess:2011yx}. The contours correspond to 63.8\%, 95\% and 99\% of
confidence level. }
\label{All2PlotCI}
\end{figure}

\subsubsection{Local Second Law of Thermodynamics}

The law of generation of \textit{local} entropy in a fluid on a FRW
space--time can be written as \cite{WeinbergBook,MisnerBook}

\begin{equation}\label{entropy_definition}
T \, \nabla_{\nu} s^{\nu} = \zeta (\nabla_{\nu} u^{\nu})^2 = 9H^2 \zeta,
\end{equation}

\noindent where $T$ is the temperature and $\nabla_{\nu} s^{\nu} $
is the rate of entropy production in a unit volume. With this, the
second law of the thermodynamics can be written as

\begin{equation}\label{2ndLawThermodynamics}
 T \nabla_{\nu} s^{\nu} \geq 0,
\end{equation}

\noindent so, from the expression (\ref{entropy_definition}), it
simply implies that $\zeta \geq 0$.

%\begin{equation}\label{2ndLawThermodynamicsZeta}
%\zeta \geq 0 
%\end{equation}

For the present model this inequality becomes (see eq.
[\ref{ViscosityDefinition}])

\begin{equation}\label{entropy_condition}
\zeta_0 \geq 0.
\end{equation}

\begin{table}
  \centering
\begin{tabular}{| c  | c c | c c |}
\multicolumn{5}{c}{\textbf{}}\\
\hline
%\multicolumn{6}{c}{Assuming a \emph{Dirac delta prior} for $H_0$} \\

%\hline
Probe &   $\zeta_0$ & $\gamma_{\rm de}$ &
$\chi^2_{{\rm min}}$ & $\chi^2_{{\rm d.o.f.}}$ \\
\hline

SNe Ia & $-0.102 \pm 0.022$  & $-0.904 \pm 0.11$   & $590.69 $   & 1.06  \\
$(R, l_A, z_*)$ CMB & $-0.0025 \pm 0.001$  & $-0.226 \pm 0.04$   &  1.01  &
1.01 \\
$H(z)$  &  $-0.003 \pm 0.022$ & $-0.199^{+0.14}_{-0.16}$   & $8.049$  &
$0.731$ \\
SNe Ia + CMB + BAO + $H(z)$  & $-0.00035 \pm 0.0007$ & $-0.363 \pm 0.028$ &
635.78 & 1.06 \\
\hline

\end{tabular}
\caption{Best estimated values of the dimensionless
coefficients  $\zeta_0$ and $\gamma_{\rm de}$.
Figures \ref{PlotCISNeU2CMB3pHz}  and \ref{All2PlotCI} show
the confidence intervals.} 
\label{TableBestEstimated}
\end{table}

%\section{Discussion}

\section{Conclusions}
\label{SectionConclusions}

It has been studied a cosmological model composed of a bulk viscous
matter fluid interacting with a dark energy fluid. 
The model is compared with cosmological observations to estimate and
constrain the values of the bulk viscous coefficient $\zeta_0$ proportional
to the Hubble parameter, and the barotropic index of the dark energy
$\gamma_{\rm de}$.
It is also used the local second law of thermodynamics (LSLT), that
states $\zeta_0 > 0$, as a criterion for the allowed values for 
$\zeta_0$.

It is found that using the combined SNe + CMB + BAO + $H(z)$ data sets,
the best estimated value of $\zeta_0$ is \textit{negative} (implying a
violation of the LSLT) and very close to zero. The confidence intervals
constrain the values of $\zeta_0$ to be very around to zero, $-0.003 <
\zeta_0 < 0.0025$, with a 99\% of confidence level. We interpret these
results as an indication that the cosmological data prefer a model with
a practically null bulk viscosity.
Since in the present model the interacting term is proportional to the bulk
viscosity, this implies also a negligible interaction between the dark
components.

On the other hand, it is found negatives values of $\gamma_{\rm de}$
with a 99\% of confidence level, corresponding to a phantom dark energy.

It may be an indicative of the phantom energy as a
preferred mechanism by the cosmological observations (in combination
with the LSLT) to explain the accelerated expansion of the Universe,
instead of the bulk viscous mechanism.

\bibliographystyle{aipproc}   % if natbib is available

\bibliography{3Proceeding}

\begin{thebibliography}{33}
\expandafter\ifx\csname natexlab\endcsname\relax\def\natexlab#1{#1}\fi
\providecommand{\enquote}[1]{``#1''}
\expandafter\ifx\csname url\endcsname\relax
  \def\url#1{\texttt{#1}}\fi
\expandafter\ifx\csname urlprefix\endcsname\relax\def\urlprefix{URL }\fi
\providecommand{\eprint}[2][]{\url{#2}}

\bibitem[Riess et~al.(1998)]{Riess:1998cb}
A.~G. Riess, et~al., \emph{Astron.J.} \textbf{116}, 1009--1038 (1998),
  \eprint{astro-ph/9805201}.

\bibitem[Perlmutter et~al.(1999)]{Perlmutter:1998np}
S.~Perlmutter, et~al., \emph{Astrophys.J.} \textbf{517}, 565--586 (1999), the
  Supernova Cosmology Project, \eprint{astro-ph/9812133}.

\bibitem[Amanullah et~al.(2010)]{Amanullah:2010vv}
R.~Amanullah, C.~Lidman, D.~Rubin, G.~Aldering, P.~Astier, et~al.,
  \emph{Astrophys.J.} \textbf{716}, 712--738 (2010), \eprint{1004.1711}.

\bibitem[Weinberg(1989)]{WeinbergCosmoConstProblem}
S.~Weinberg, \emph{Rev. Mod. Phys.} \textbf{61}, 1--23 (1989),
  \urlprefix\url{http://link.aps.org/doi/10.1103/RevModPhys.61.1}.

\bibitem[Padmanabhan(2003)]{Padmanabhan:2002ji}
T.~Padmanabhan, \emph{Phys.Rept.} \textbf{380}, 235--320 (2003),
  \eprint{hep-th/0212290}.

\bibitem[Carroll(2001)]{Carroll:2000fy}
S.~M. Carroll, \emph{Living Rev.Rel.} \textbf{4}, 1 (2001),
  \eprint{astro-ph/0004075}.

\bibitem[Steinhardt et~al.(1999)]{Steinhardt:1999nw}
P.~J. Steinhardt, L.-M. Wang, and I.~Zlatev, \emph{Phys.Rev.} \textbf{D59},
  123504 (1999), \eprint{astro-ph/9812313}.

\bibitem[Chimento et~al.(2000)]{ChimentoJakubiPavon2000}
L.~P. Chimento, A.~S. Jakubi, and D.~Pav{\'o}n, \emph{Phys. Rev. D}
  \textbf{62}, 063508 (2000),
  \urlprefix\url{http://link.aps.org/doi/10.1103/PhysRevD.62.063508}.

\bibitem[Kremer and Sobreiro(2011)]{Kremer:2011cd}
G.~M. Kremer, and O.~A. Sobreiro  (2011), \eprint{1109.5068}.

\bibitem[Heller and Klimek(1975)]{HellerKlimek1975}
M.~Heller, and Z.~Klimek, \emph{Astrophysics and Space Science} \textbf{33},
  L37--L39 (1975).

\bibitem[Barrow(1986)]{Barrow1986}
J.~D. Barrow, \emph{Physics Letters B} \textbf{180}, 335--339 (1986), ISSN
  0370-2693,
  \urlprefix\url{http://www.sciencedirect.com/science/article/pii/037026938691%
1986}.

\bibitem[Padmanabhan and Chitre(1987)]{Visco-Padmanabhan1987}
T.~Padmanabhan, and S.~Chitre, \emph{Physics Letters A} \textbf{120}, 433--436
  (1987), ISSN 0375-9601,
  \urlprefix\url{http://www.sciencedirect.com/science/article/pii/037596018790%
1046}.

\bibitem[Gron(1990)]{Visco-Gron1990}
O.~Gron, \emph{Astrophys. Space Sci.} \textbf{173}, 191--225 (1990).

\bibitem[Maartens(1995)]{Visco-Maartens1995}
R.~Maartens, \emph{Classical and Quantum Gravity} \textbf{12}, 1455 (1995),
  \urlprefix\url{http://stacks.iop.org/0264-9381/12/i=6/a=011}.

\bibitem[Zimdahl(1996)]{Visco-Zimdahl1996}
W.~Zimdahl, \emph{Phys. Rev. D} \textbf{53}, 5483--5493 (1996),
  \urlprefix\url{http://link.aps.org/doi/10.1103/PhysRevD.53.5483}.

\bibitem[Cataldo et~al.(2005)]{Cataldo:2005qh}
M.~Cataldo, N.~Cruz, and S.~Lepe, \emph{Phys.Lett.} \textbf{B619}, 5--10
  (2005), \eprint{hep-th/0506153}.

\bibitem[Colistete et~al.(2007)]{Colistete:2007xi}
R.~Colistete, J.~Fabris, J.~Tossa, and W.~Zimdahl, \emph{Phys.Rev.}
  \textbf{D76}, 103516 (2007), \eprint{0706.4086}.

\bibitem[Avelino and Nucamendi(2009)]{Avelino:2008ph}
A.~Avelino, and U.~Nucamendi, \emph{JCAP} \textbf{0904}, 006 (2009),
  \eprint{0811.3253}.

\bibitem[Avelino and Nucamendi(2010)]{Avelino:2010pb}
A.~Avelino, and U.~Nucamendi, \emph{JCAP} \textbf{1008}, 009 (2010),
  \eprint{1002.3605}.

\bibitem[Hip{\'o}lito-Ricaldi et~al.(2010)]{Visco-RicaldiVeltenZimdahl2010}
W.~S. Hip{\'o}lito-Ricaldi, H.~E.~S. Velten, and W.~Zimdahl, \emph{Phys. Rev.
  D} \textbf{82}, 063507 (2010),
  \urlprefix\url{http://link.aps.org/doi/10.1103/PhysRevD.82.063507}.

\bibitem[Montiel and Breton(2011)]{Visco-AMontielNBreton2011}
A.~Montiel, and N.~Breton, \emph{JCAP} \textbf{1108}, 023 (2011),
  \eprint{1107.0271}.

\bibitem[Zimdahl(2000)]{Zimdahl:1999tn}
W.~Zimdahl, \emph{Phys.Rev.} \textbf{D61}, 083511 (2000),
  \eprint{astro-ph/9910483}.

\bibitem[Mathews et~al.(2008)]{Mathews:2008hk}
G.~Mathews, N.~Lan, and C.~Kolda, \emph{Phys.Rev.} \textbf{D78}, 043525 (2008),
  \eprint{0801.0853}.

\bibitem[Chimento et~al.(2009)]{Chimento:2007yt}
L.~P. Chimento, M.~I. Forte, and G.~M. Kremer, \emph{Gen.Rel.Grav.}
  \textbf{41}, 1125--1137 (2009), \eprint{0711.2646}.

\bibitem[Komatsu et~al.(2011)]{Komatsu:2010fb}
E.~Komatsu, et~al., \emph{Astrophys.J.Suppl.} \textbf{192}, 18 (2011),
  \eprint{1001.4538}.

\bibitem[Hu and Sugiyama(1996)]{Hu:1995en}
W.~Hu, and N.~Sugiyama, \emph{Astrophys.J.} \textbf{471}, 542--570 (1996),
  revised version, \eprint{astro-ph/9510117}.

\bibitem[Percival et~al.(2010)]{Percival:2009xn}
W.~J. Percival, et~al., \emph{Mon.Not.Roy.Astron.Soc.} \textbf{401}, 2148--2168
  (2010), 21 pages, 15 figures, \eprint{0907.1660}.

\bibitem[Eisenstein and Hu(1998)]{Eisenstein:1997ik}
D.~J. Eisenstein, and W.~Hu, \emph{Astrophys.J.} \textbf{496}, 605 (1998),
  \eprint{astro-ph/9709112}.

\bibitem[Stern et~al.(2010)]{Stern:2009ep}
D.~Stern, R.~Jimenez, L.~Verde, M.~Kamionkowski, and S.~A. Stanford,
  \emph{JCAP} \textbf{1002}, 008 (2010), \eprint{0907.3149}.

\bibitem[Gaztanaga et~al.(2009)]{Gaztanaga:2008xz}
E.~Gaztanaga, A.~Cabre, and L.~Hui, \emph{Mon.Not.Roy.Astron.Soc.}
  \textbf{399}, 1663--1680 (2009), \eprint{0807.3551}.

\bibitem[Riess et~al.(2011)]{Riess:2011yx}
A.~G. Riess, L.~Macri, S.~Casertano, H.~Lampeitl, H.~C. Ferguson, et~al.,
  \emph{Astrophys.J.} \textbf{730}, 119 (2011), \eprint{1103.2976}.

\bibitem[Weinberg(1972)]{WeinbergBook}
S.~Weinberg, \emph{Gravitation and Cosmology: principles and applications of
  the general theory of relativity}, John Wiley \& Sons Inc, New York, USA,
  1972.

\bibitem[Misner and Wheeler(1973)]{MisnerBook}
C.~W. Misner, and J.~A. Wheeler, \emph{Gravitation}, W. H. Freeman and Company,
  USA, 1973.

\end{thebibliography}

\IfFileExists{\jobname.bbl}{}
 {\typeout{}
  \typeout{******************************************}
  \typeout{** Please run "bibtex \jobname" to optain}
  \typeout{** the bibliography and then re-run LaTeX}
  \typeout{** twice to fix the references!}
  \typeout{******************************************}
  \typeout{}
 }

\end{document}